\documentclass{epl}

\usepackage{graphicx,psfrag}

\title{Electronic Raman Spectra of Superconducting Borocarbides}
\author{
   H. Jang\inst{1} \and H. Won\inst{1} \and K. Maki\inst{2,3}}
\institute{
  \inst{1}Department of Physics, Hallym University,
Chunchon, 200-702, South Korea\\
  \inst{2}Max-Planck-Institute for the Physics of Complex Systems,
N\"othnitzer Str.38, 01187 Dresden, Germany\\
  \inst{3}Department of Physics and Astronomy, University of Southern
California, Los Angeles, CA 90089-0484, USA
}

\begin{document}

\maketitle

\begin{abstract}
Recently
we have proposed a $s+g$-wave model for the
superconductivity in borocarbides,
YNi$_2$B$_2$C and LuNi$_2$B$_2$C \cite{Maki_0110591,Izawa-prl2002,Thalmeier-acta2003}.
In the present paper we first summarize thermodynamic
properties of $s+g$-wave model.
Then we shall analyse the recent Raman spectra data of
RNi$_2$B$_2$C (with R=Lu and Y) by Yang {\it et al.}\cite{yang}.
The present model appears to describe salient features of the Raman spectra.

\end{abstract}

\section{Introduction}
The Superconductivity in rare earth borocarbides is of  great
interest \cite{canfield,muller}. In particular its interplay with magnetism
and superconductivity is very  fascinating\cite{amici}. However in the
following we limit ourselves to superconducting borocarbides
LuNi$_2$B$_2$C and YNi$_2$B$_2$C. They have relatively high
superconducting transition temperature $T_c=16.5$ K and 15.5 K,
respectively. Although the dominance of $s$-wave component in
$\Delta({\bf k})$ has been established by substituting Ni by Pt and
subsequent opening  of the energy gap \cite{nohara,borkowski}, the
superconductivity exhibits a number of peculiarities common to
nodal superconductivities\cite{won_muller,maki_fluc}. For example both the
$\sqrt{H}$ dependence of the specific heat and the $H$
linear dependence of the thermal conductivity indicate that the
superconductivity has the nodal
excitations\cite{volovik,Nohara97,freudenberger,boaknin,dahm,euro01,won-current} similar to $d$-wave
superconductors in high $T_c$ cuprate superconductors. Further the
presence of de Haas van Alphen oscillation in the vortex state of
LuNi$_2$B$_2$C down
to $H=0.2H_{c2}$ indicates again the nodal superconductors
\cite{terashima}. In conventional $s$-wave superconductor de Haas van
Alphen oscillation would disappear for $H<0.8H_c$ \cite{maki_prb}.
Further the upper critical field determined for LuNi$_2$B$_2$C and
YNi$_2$B$_2$C \cite{Metlushko} in a magnetic field within the a-b plane
exhibits clear fourfold symmetry reminiscent to $d$-wave
superconductors\cite{Wang}.
\par
These experiments indicate clearly $\Delta({\bf k})$ in
borocarbides has to have an anisotropic $s$-wave order parameter.
Further a) $\Delta({\bf k})$ has to have the nodal structure or the
quasiparticle density of states, $N(E) \sim |E|$ for $|E| \ll \Delta$
where $\Delta$ is the superconducting order parameter (i.e. the maximum of
$\Delta({\bf k})$), which gives both
the $\sqrt{H}$dependence of the specific heat and the $H$ linear
thermal conductivity in the vortex state. b) the nodal structure
has to have the fourfold symmetry within the $a$-$b$ plane and
to be consistent with the tetragonal symmetry. These two constraints appear
to suggest almost uniquely\cite{Maki_0110591,Izawa-prl2002}
% ----------------------------------------  1½Ä
\begin{equation}
\Delta({\bf k})~ =~ \frac12\Delta(1 ~- \sin^4\theta\cos(4\phi))
\end{equation}
or $s$+$g$-wave superconductor.
Here $\theta$ and $\phi$ are the polar and azimuthal angles describing
${\bf k}$.
Contrary to Ref.\cite{Maki_0110591}
we have minus(-) sign in front of the $g$-wave term.
This corresponds to point nodes at [100], [010], etc.
Those positions of the nodal points are consistent with
the magnetothermal conductivity data\cite{Izawa-prl2002}.
More recently the point node at the same positions
have been seen by the magnetospecific heat measurement
in the vortex state of YNi$_2$B$_2$C  by Park {\it et al.}\cite{park}.
We shall see later the Raman spectra data from both
YNi$_2$B$_2$C and LuNi$_2$B$_2$C
are consistent with Eq.(1).
Recently
the above
Raman data have been analysed in terms of 2D $s+g$-wave
model by Lee and Choi\cite{hclee}.
Unlike the present model their model has line nodes.
Therefore their model cannot describe the magnetothermal conductivity data\cite{Izawa-prl2002}.
Further the description of the Raman spectra within this model is unsatisfactory.

\begin{figure}
%\vspace{-60mm}
\twofigures[scale=0.4]{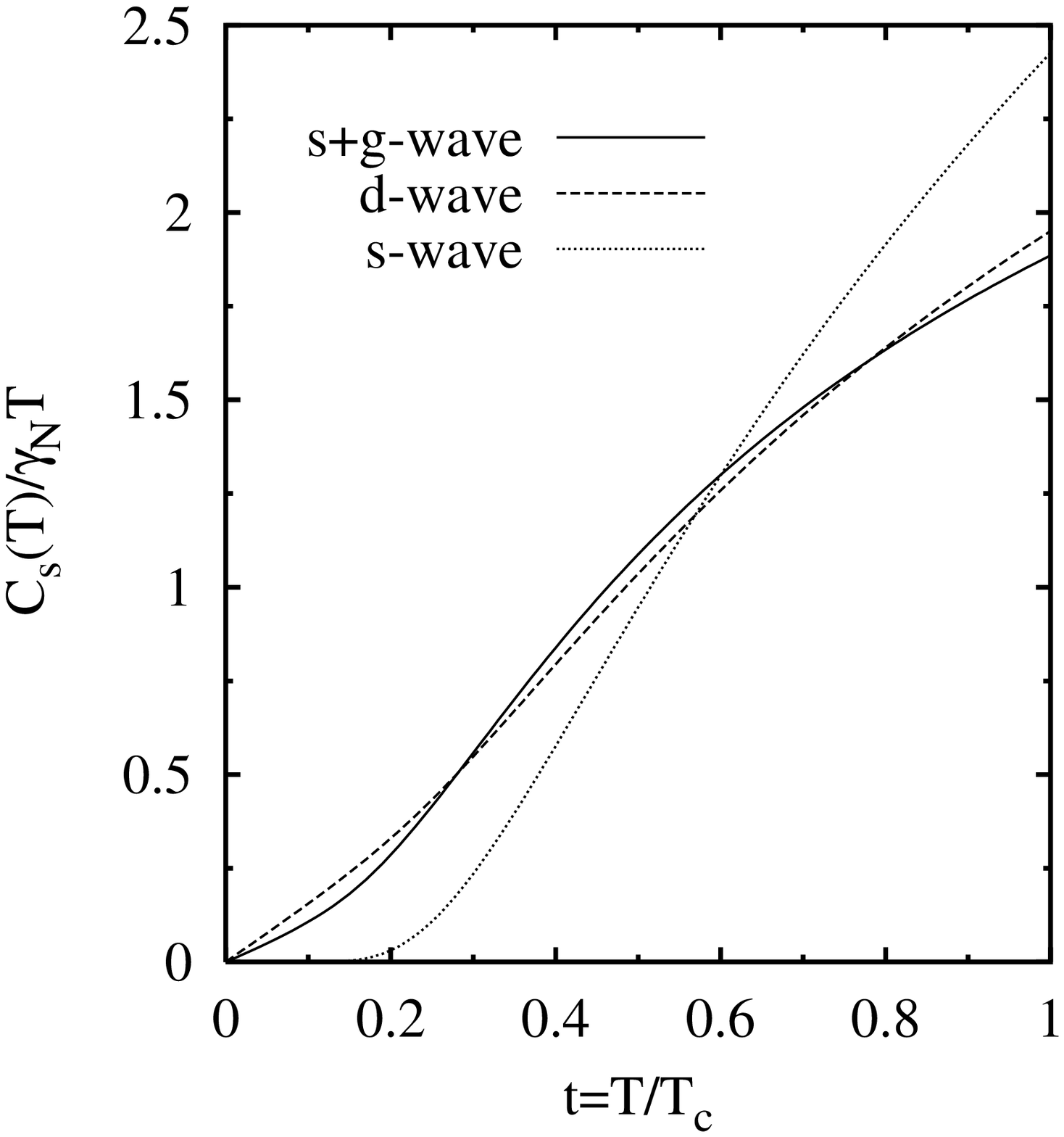}{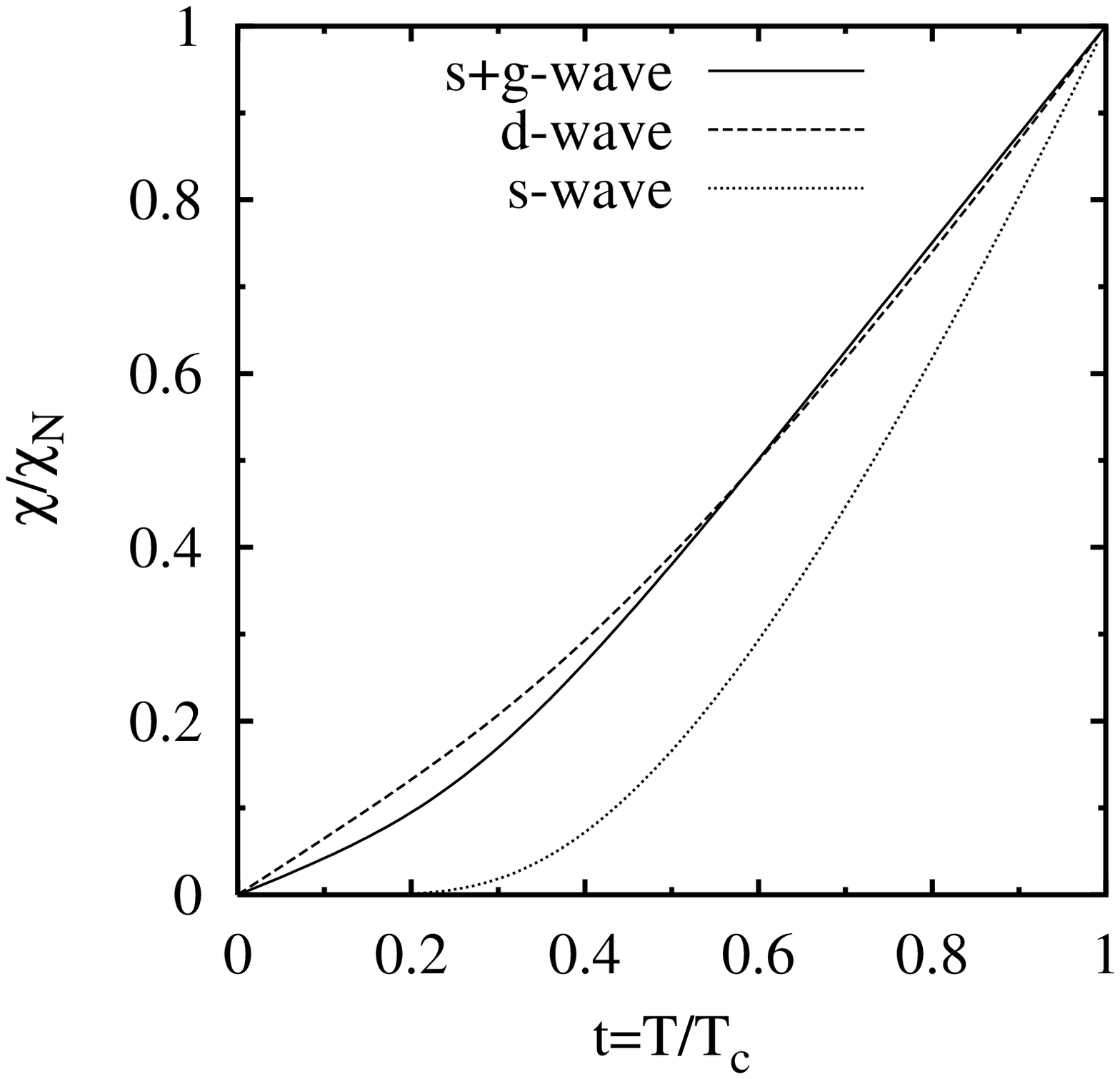}
\caption{The specific heats normalized by the normal state one
are
shown for $s+g$-wave, $s$-wave and $d$-wave superconductors.
$\gamma_N$ is the Sommerfeld Constant.
}
\caption{The spin susceptibility normalized by the normal state
are
shown for $s+g$-wave, $s$-wave and $d$-wave superconductors.}
\end{figure}

%\vspace{-3.0cm}
%\begin{figure}
%\includegraphics[width=80mm]{fig2.ps}
%\caption{
%Normalized quasiparticle density of states for $s+g$ model.
%}
%\vspace{2cm}
%\end{figure}

\section{Thermodynamics}
First we determine the temperature dependent energy gap $\Delta(T)$
within the weak coupling theory.
Here $\Delta(T)$ is the maximum value of the energy gap
$\Delta({\bf k})$.
\begin{equation}
\lambda^{-1}=<f^2>^{-1} \int_0 ^{E_c} \!\!\!\! d E \,\,\big< {\rm Re}
\frac{f^2}{\sqrt{E^2-\Delta^2 f^2 }}>\tanh(\frac{E}{2T})
\end{equation}
where $f=\frac12(1- \sin^4\theta\cos(4\phi))$.
$\lambda$ and $E_c$ are the dimensionless coupling constant and
the cut-off energy($\gg \Delta $), respectively.
Further $< \ldots> = \int {d \Omega}/{4\pi}$.
\par
In the vicinity of $T=T_c$, Eq.(2) gives
\begin{equation}
\Delta^2(T) \simeq
\frac{2(2\pi T)^2}{7\zeta(3)} \frac{<f^2>}{<f^4>} (-\ln \frac{T}{T_c})
\end{equation}
with $T_c=\frac{2\gamma}{\pi}E_c e^{-1/\lambda}$
and $\gamma=1.78107\ldots$ the Euler constant.
On the other hand for $T \ll \Delta_0$
\begin{eqnarray}
-\ln(\frac{\Delta(T)}{\Delta_0})
&=& <f^2>^{-1} \big\{\frac{3\pi}{16}\zeta(3)(\frac{T}{\Delta})^3
+\frac{7\pi^4}{160}(\frac{T}{\Delta})^4 + \ldots  \big\}\\
&=&1.9635(\frac{T}{\Delta})^3 + 14.2 (\frac{T}{\Delta})^4 \ldots
\end{eqnarray}
where
\begin{equation}
\Delta_0=\Delta(0) =
\frac{2\gamma}{\pi}
T_c \exp[-<f^2>^{-1}<f^2 \ln |f|>] \simeq 2.76 T_c
\end{equation}
Then the specific heat $C_s$ is given by
\begin{eqnarray}
C_s &=&  T^{-2}N_0 \int_0^{\infty} d \xi
< {\rm sech}^2(\frac{E}{2T})\,\, (E^2 - \frac{T}{2} \frac{d \Delta^2}{d T}f^2) > \\
&\simeq &\frac{2\pi^2 }{3}  N_0 T\big\{
\frac{27}{4\pi}\zeta(3)(\frac{T}{\Delta}) ~+~
\frac{63}{80}(\frac{ T}{\Delta})^2 ~+ \ldots \big\}
\end{eqnarray}
where
$E=\sqrt{\xi^2 + \Delta^2 f^2}$ and $N_0$ is the density of states
in the normal state.
In Fig.1 we show $C_s/(\frac{2\pi^2 }{3}  N_0 T)$ versus $T/T_c$
for $s+g$-wave, $s$-wave and $d$-wave superconductors.
As is readily seen the specific heat of $s+g$-wave
superconductor is very similar to the one in $d$-wave
superconductors\cite{Won-prb94}.
Also the spin susceptibility of $s+g$-wave superconductor
is shown in Fig.2, which is very similar to the one in d-wave
superconductors.
\par
The superfluid density, on the other hand, has the axial symmetry.
The superfluid density in the a-b plane is given by
\begin{eqnarray}
\frac{\rho_{s,ab}(T)}{\rho_{s,ab}(0)} &=& 1-\frac{3}{4T}  \int_0^{\infty}\!\!\!\! d \xi
<\sin^2\theta \,\,\,{\rm sech}^2(\frac{E}{2T})>  \\
&\simeq &  1-\frac{3\pi}{4}(\ln 2)(\frac{T}{\Delta}) -
\frac{5\pi^2}{32}(\frac{T}{\Delta})^2 + \ldots
\end{eqnarray}
while the superfluid density parallel to the $c$-axis is given by
\begin{eqnarray}
\frac{\rho_{s,c}(T)}{\rho_{s,c}(0)} &=& 1-\frac{3}{2T}  \int_0^{\infty} d \xi
<\cos^2\theta\,\,\, {\rm sech}^2(\frac{E}{2T})>  \\
&\simeq &  1-\frac{\pi^2}{4}(\frac{T}{\Delta})^2 -
\frac{783\pi}{256}(\frac{T}{\Delta})^3 + \ldots
\end{eqnarray}

\begin{figure}
\vspace{-40mm}
\twofigures[scale=0.4]{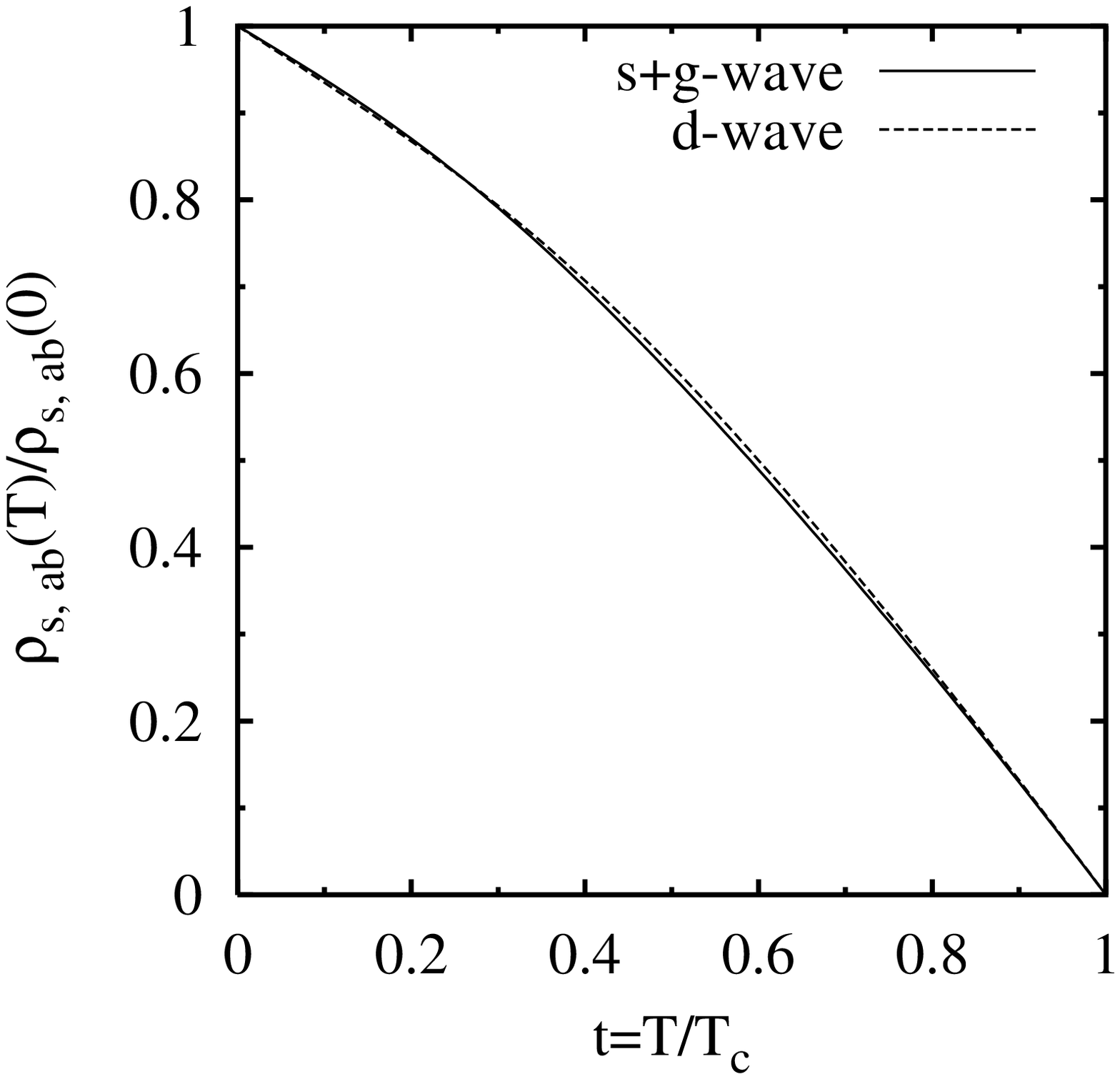}{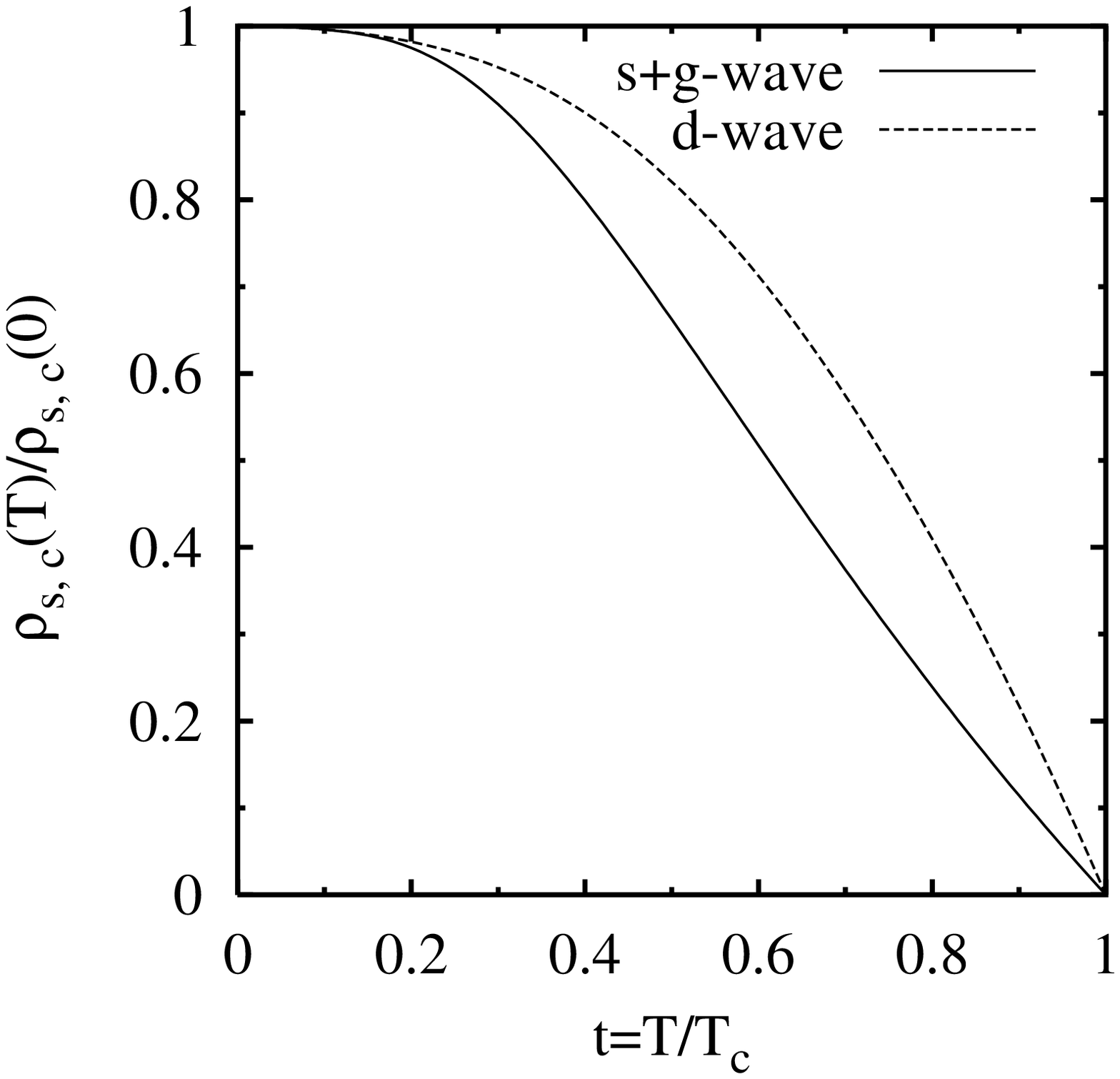}
\caption{
the superfluid density in the a-b plane for $s+g$- and $d$-wave superconductors.
}
\caption{the superfluid density parallel to the c-axis for for $s+g$- and $d$-wave superconductors. }
\end{figure}

In Fig.3 and Fig.4 we show the superfluid density in the a-b plane
and parallel to the c-axis, respectively.
The superfluid density in the a-b plane is very similar to the one in
$d$-wave superconductor.
Further the superfluid density parallel to the c-axis is somewhat
similar to the one in the $d$-wave superconductor which is due the
coherent Josephson tunneling\cite{won-klwer2001}.
We note also $T^{-1}_{1} \sim T^3$ behavior has been already
observed in Ref.\cite{Zheng}.

\par
Further the thermal conductivity within the a-b plane exhibits
the universal heat conduction \cite{Sun,lee}
% ----------------------------------------  7½Ä
\begin{equation}
\kappa/T ~=~ \frac{\pi^2}{8}\frac{n}{\Delta_0 m}
\end{equation}
in the limit
T$\rightarrow 0$ K.
Also the thermal conductivity for ${\bf H} \parallel {\bf c}$
gives
\begin{equation}
\kappa(H)/\kappa_n ~=~
\frac{3}{\pi}\frac{{v_a^2}(eH)}{\Delta^2}
\end{equation}
for $T, \sqrt{\Gamma\Delta} \ll v_a \sqrt{eH}$,
where $\Gamma$ and $v_a$ are the electron scattering rate
and the Fermi velocity within the a-b plane, respectively.  $\kappa_n$ is the one in the normal state and $v_a$ is the
Fermi velocity within the a-b plane.
Indeed the $H$-linear thermal conductivity is observed recently by
Boaknin {\it et al.}\cite{boaknin}.
\par
Also very recently the specific heat of the vortex state in
YNi$_2$B$_2$C in a magnetic field within the a-b plane is observed by Park
{\it et al.}\cite{park}.
It exhibits cusps at ${\bf H} \parallel {\bf a}$ and
${\bf H} \parallel {\bf b}$
typical to the point nodes in $s+g$-wave superconductors.

\begin{figure}
\onefigure[angle=-90,scale=0.6]{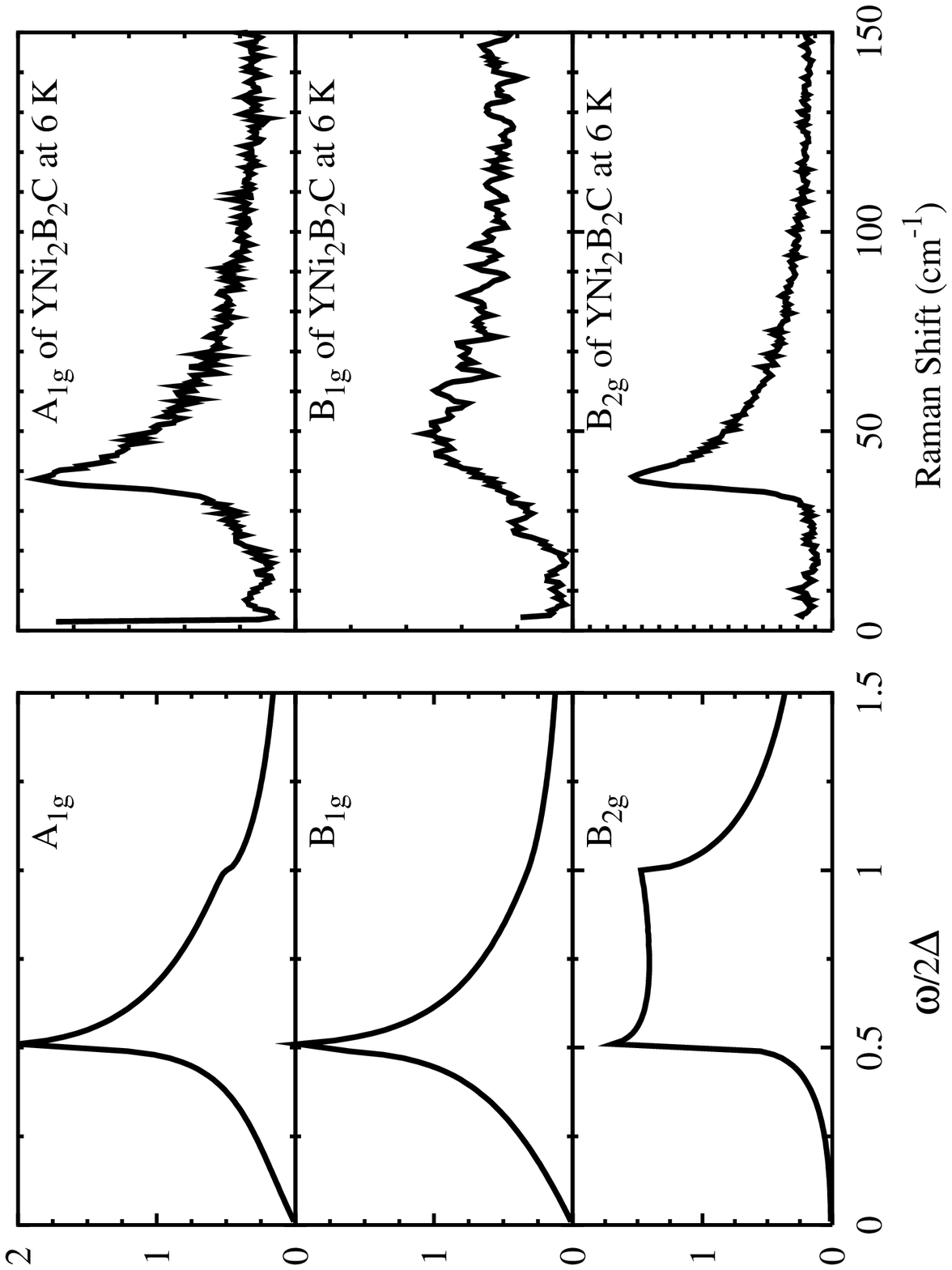}
\caption{
The theoretical Raman spectra in $A_{1g}$, $B_{1g}$ and $B_{2g}$  modes
using $s+g$ model at $T=0$ K (left panel) and  the experimental Raman spectra
for
YNi$_2$B$_2$C
taken at $T=6$ K (right panel) are shown.
}
\end{figure}

\section{Electronic Raman Spectra}
We consider the case the polarization vector of photons lie in the
$a$-$b$ plane.
Then the Raman spectra are given by \cite{Devereaux_prl,Devereaux_prb}
\begin{equation}
S_i(\omega/2\Delta) ~=~ {\rm Im}\big[<\gamma_i^2\lambda> ~-~
\frac{<\gamma_i\lambda>^2}{<\lambda>}\big]
\end{equation}
where
$\gamma_{A1g}=\sqrt{2}\cos(4\phi_k)$, $\gamma_{B1g}=\sqrt{2}\cos(2\phi_k)$
and $\gamma_{B2g}=\sqrt{2}\sin(2\phi_k)$ and
% ----------------------------------------  10=D
\begin{eqnarray}
\lambda &=& \lambda' ~+~ i\lambda'' \\ \nonumber
 \lambda' ~&=&
 \frac{f^2}{x\sqrt{f^2-x^2}}\tan^{-1}(\frac{x}{\sqrt{f^2-x^2}})\theta(f^2-x^2)\\
 &-&
 \frac{f^2}{x\sqrt{x^2-f^2}}\tanh^{-1}(\frac{\sqrt{x^2-f^2}}{x})
\theta(x^2-f^2)\\
 \lambda'' ~&=& \frac{\pi}{2x}\frac{f^2}{\sqrt{x^2-f^2}}\theta(x^2-f^2)
\end{eqnarray}
where $x=\omega/2\Delta$ and $f=\Delta({\bf k})/\Delta$ and
$< \ldots >$ means the ${\bf k}$(angle) average for the Fermi surface.
$\theta(x)$ is the step function, i.e., $\theta(x) =1$ for $x>0$ and
$\theta(x)=0$ for $x<0$.
We note for the present model the second term in Eq.(10) does not
contribute for $B_{1g}$ and $B_{2g}$ modes
due to the symmetry constraints. The electronic Raman spectra from these 3 modes are
obtained numerically and shown in Fig.3 (the left-side panel).
In parallel to the theoretical result we show the experimental data for
YNi$_2$B$_2$C
taken at $T=6$ K.

\par
We note $A_{1g}$ mode is very consistent with the observed spectra.
Also the low frequency parts ($\omega \le \Delta(T)$) of both
the $B_{1g}$ mode and the $B_{2g}$ mode are very consistent.
On the other hand the theoretical curve for the $B_{2g}$
mode exhibits a cusp at $\omega = 2\Delta(T)$, which is not seen
experimentally.
We don't know if the cusp-like feature will disappear
with increasing temperature or not.
Also the peak position of the $B_{1g}$ mode is somewhat in
the higher energy than that of the $B_{2g}$ mode.
Again we don't know if this is the effect of the temperature.
In any event we may conclude that $s+g$-wave model captures
the main feature of the Raman spectra of YNi$_2$B$_2$C.
Also the peak position in the $B_{2g}$ mode give $\Delta(T)$.
\par
The weak coupling theory gives $\Delta_0$ for
YNi$_2$B$_2$C($T_c \simeq 15.3$ K) and
LuNi$_2$B$_2$C ($T_c \simeq 15.7$ K)
is 42.2 K(3.64 meV) and 43.3 K(3.73 meV), respectively.
On the other hand the data at $T/T_c \simeq 1/2$ indicate
$\Delta_0$=50.4 K and 64.7 K for YNi$_2$B$_2$C and
LuNi$_2$B$_2$C, respectively.
Therefore we may conclude that the borocarbides superconductors are
in the intermediate coupling region.
However, clearly a further experiment at low temperatures is highly
desirable.

\section{Summary}
We have analysed further the $s+g$-wave superconductors proposed in
\cite{Maki_0110591,Izawa-prl2002}.
This model appears to describe the recent specific heat data \cite{park} as
well.
We have also studied the Raman spectra reported in Ref.\cite{yang}.
The present model appears to capture the main feature of the observed spectra.
This further confirms the presence of the order parameter
$\Delta({\bf k})$ with point nodes at ${\bf k}$ =(100), (010), ($\bar{1}$00),
and (0$\bar{1}$0).

\acknowledgements
We would like to thank Peter Thalmeier for useful
discussions on $s+g$ model, and In-Sang Yang for sending
experimental Raman data which we show in the right hand panel of Fig.5.
K.M. also thanks the
hospitality of Department of Physics at Hallym University where a part
of this work was done.
H.W. thanks the hospitality of MPIPKS, Dresden for the completion
of this work.
This research was supported by Hallym University through the 2002 research fund.

\end{document}